\documentclass[sigconf]{acmart}

\def\BibTeX{{\rm B\kern-.05em{\sc i\kern-.025em b}\kern-.08emT\kern-.1667em\lower.7ex\hbox{E}\kern-.125emX}}
\copyrightyear{2019}
\acmYear{2019}
\setcopyright{acmlicensed}

\usepackage[]{amsmath}
\usepackage{breqn}
\usepackage{array}
\usepackage{fixltx2e}
\usepackage{url}
\usepackage{graphics}
\usepackage{graphicx}
\usepackage{subfig}
\usepackage{float}
\usepackage{amssymb}
\usepackage{multirow}
\usepackage{tabularx,multicol,multirow}
\usepackage[english]{babel}
\usepackage[utf8]{inputenc}
\usepackage{algorithmicx}
\usepackage{algorithm}
\usepackage{algpseudocode}
\usepackage{footmisc}

\algnewcommand\algorithmicinput{\textbf{Input:}}
\algnewcommand\INPUT{\item[\algorithmicinput]}
\algnewcommand\algorithmicoutput{\textbf{Output:}}
\algnewcommand\OUTPUT{\item[\algorithmicoutput]}
\newcolumntype{C}{>{\centering\arraybackslash}X}
\newcolumntype{P}[1]{>{\centering\arraybackslash}p{#1}}

\begin{document}

%
\title{PUNCH: Positive UNlabelled Classification based information retrieval in Hyperspectral images}

%
\author{Anirban Santara}
\authornote{Both authors contributed equally to this research.}
\email{anirban\_santara@iitkgp.ac.in}
\orcid{0002-1571-3885}
\affiliation{%
  \institution{Indian Institute of Technology, Kharagpur}
  \city{Kharagpur}
  \state{WB}
  \country{India}
  \postcode{721302}
}

\author{Jayeeta Datta}
\authornotemark[1]
\email{djayeeta@seas.upenn.edu}
\affiliation{%
  \institution{University of Pennsylvania}
  \streetaddress{3330 Walnut Street}
  \city{Philadelphia}
  \state{PA}
  \postcode{19104-6309}
}

\author{Sourav Sarkar}
\email{ss5645@columbia.edu}
\affiliation{%
  \institution{Columbia University}
  \streetaddress{500 W 120th St}
  \city{New York}
  \state{NY}
  \postcode{10027}
}

\author{Ankur Garg}
\email{agarg@sac.isro.gov.in}
\affiliation{%
  \institution{Space Applications Centre, Indian Space Research Organization (ISRO)}
  \city{Ahmedabad}
  \state{GJ}
  \postcode{380015}
  \country{India}
}
 
\author{Kirti Padia}
\email{kirtipadia@sac.isro.gov.in}
\affiliation{%
  \institution{Space Applications Centre, Indian Space Research Organization (ISRO)}
  \city{Ahmedabad}
  \state{GJ}
  \postcode{380015}
  \country{India}
}

\author{Pabitra Mitra}
\email{pabitra@cse.iitkgp.ernet.in}
\affiliation{%
  \institution{Indian Institute of Technology, Kharagpur}
  \city{Kharagpur}
  \state{WB}
  \country{India}
  \postcode{721302}
}

%

\renewcommand{\shortauthors}{Santara and Datta, et al.}

%
\begin{abstract}
Hyperspectral images of land-cover captured by airborne or satellite-mounted sensors provide a rich source of information about the chemical composition of the materials present in a given place. This makes hyperspectral imaging an important tool for earth sciences, land-cover studies, and military and strategic applications. However, the scarcity of labeled training examples and spatial variability of spectral signature are two of the biggest challenges faced by hyperspectral image classification. In order to address these issues, we aim to develop a framework for material-agnostic information retrieval in hyperspectral images based on Positive-Unlabelled (PU) classification. Given a hyperspectral scene, the user labels some positive samples of a material he/she is looking for and our goal is to retrieve all the remaining instances of the query material in the scene. Additionally, we require the system to work equally well for any material in any scene without the user having to disclose the identity of the query material. This material-agnostic nature of the framework provides it with superior generalization abilities. We explore two alternative approaches to solve the hyperspectral image classification problem within this framework. The first approach is an adaptation of non-negative risk estimation based PU learning for hyperspectral data. The second approach is based on one-versus-all positive-negative classification where the negative class is approximately sampled using a novel spectral-spatial retrieval model. We propose two annotator models -- uniform and blob -- that represent the the labelling patterns of a human annotator. We compare the performances of the proposed algorithms for each annotator model on three benchmark hyperspectral image datasets -- Indian Pines, Pavia University and Salinas.

\end{abstract}

%
%


%

%

%
\maketitle

\section{Introduction}
\label{sec:intro}
\begin{figure*}
  \centering
  \includegraphics[width=0.9\textwidth]{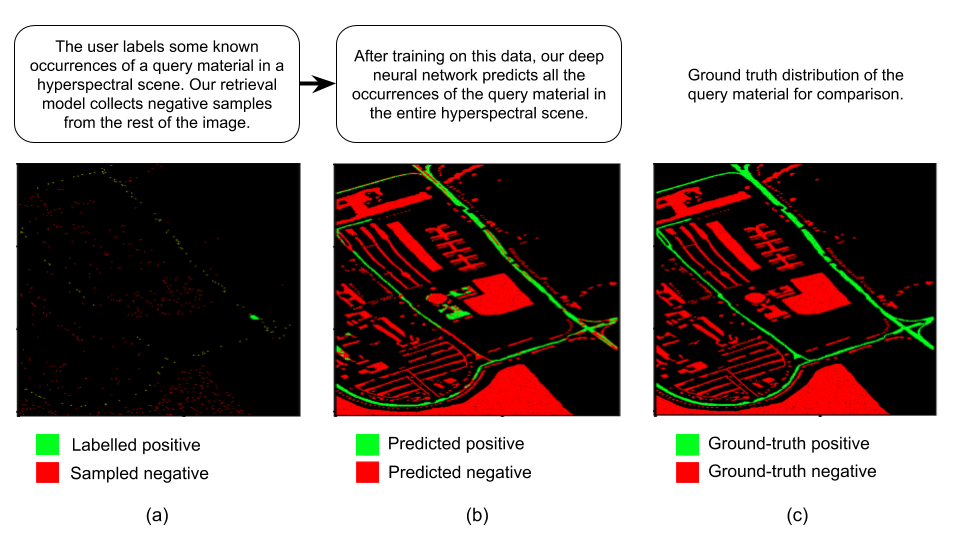}
  \caption{Workflow of PUNCH. Hyperspectral scene -- Pavia University; query material -- Asphalt.}
  \Description{}
  \label{fig:teaser}
\end{figure*}
Hyperspectral imaging (HSI) \cite{Landgrebe:2002, Richards:2013} measures reflected radiation from a surface at a series of narrow, contiguous frequency bands. It differs from multi-spectral imaging which senses a few wide, separated frequency bands. Hyperspectral imaging produces three-dimensional $(x, y, \lambda)$ data volumes, where $x$ and $y$ represent the spatial dimensions and $\lambda$ represents the spectral dimension. Such detailed spectra contain fine-grained information about the chemical composition of the materials in a scene that is richer than is available from a multi-spectral image \cite{multi-vs-hyper}.\\

Majority of the existing literature on hyperspectral image classification frame it as a supervised \cite{Tuia:2013, Valls:2005, Melgani:2004, Li:2016, santara2017bass} or semi-supervised \cite{CampsValls2007SemiSupervisedGH, Buchel2018LadderNF, Cui2018SemiSupervisedCO} multi-class classification problem. Hyperspectral images pose a unique set of challenges when it comes to multi-class classification. Each material with a distinct spectral identity is a class. Hence the number of possible classes is countably infinite. When it comes to land-cover, the same species of crop (for example, wheat) can have drastically different spectral signatures depending on the location where it is grown (due to differences in chemical composition of soil and water) and the time of the year (temperature, humidity, rainfall, etc) \cite{Herold2004SpectrometryFU, Rao2007DevelopmentOA}. Hence, creating a standardized library of spectral signatures of materials considering all these factors of variability is hard. Over and above, multi-class classification requires ground-truth labels for each class. Collection of ground-truth amounts to sourcing samples of a material from the exact location and exact time of the year and recording their spectral signatures \cite{ShepherdDevelopmentOR} - a process that is impractical and intractable to be performed at scale. Also, different hyperspectral imaging systems produce images with different physical properties depending on the spectral response of the sensor, resolution, altitude, illumination and mode of capture (airborne vs. spaceborne), distortions and so on. As a result, multi-class classification models trained on open-source (but relatively old) benchmark datasets like Indian Pines and Pavia have extremely limited efficacy when it comes to large-scale deployment in real life applications.\\

In this work, we formulate the hyperspectral classification problem as one that is material-agnostic, imaging system independent and not contingent on an extensive ground-truth labelling effort. The motivation is deployment at scale. Given a hyperspectral scene, the user marks-up some known occurrences of the query material. No information is provided about pixels that do not contain the query material. The goal of the system is to locate all other occurrences of the same material in the scene with high precision and recall. The system should work for any target material with a distinct spectral-signature and it should not require the user to disclose the identity of the material being searched.\\

Our formulation builds upon the classical problem of Content-Based Information Retrieval (CBIR) in multi-media databases \cite{yoshitaka1999survey, smeulders2000content, liu2007survey}. Given a query item, the task is to retrieve items from a database that are similar in content. At the heart of CBIR lies the task of designing a \emph{retrieval model}. A retrieval model is a function that returns a score that is an estimate of the similarity of an element of the database with the query element. These scores can be used to find the most relevant elements for output. With the advent of deep learning, CBIR has witnessed unprecedented records of success in domains ranging from images \cite{Kirzhevsky:2012, wan2014deep, lin2015deep} to text \cite{mitra2017learning} and audio \cite{van2013deep} and multi-modal information retrieval \cite{kiros2014unifying, wang2016effective}.\\

We approach the problem from a Positive-Unlabelled (PU) classification \cite{Denis2000LearningFP, Zhang2008LearningFP, elkan2008learning, Hou2018GenerativeAP} perspective. PU classification algorithms are specialized to deal with the setting where the training data comprises of positive samples labelled by the user and unlabelled samples that may consist of both positive and negative classes. There are two main approaches to PU classification. The first approach is based on heuristic-driven intelligent sampling of the negative class followed by supervised training of a binary Positive-Negative (PN) classifier with the labelled positive and sampled negative examples \cite{Nigam1998LearningTC}. The second approach is based on non-negative risk-estimation in which the unlabelled data is treated as negative data with lesser weights \cite{Lee2003LearningWP}. We explore both categories of algorithms in this work and present a comparison of performance results. Deep Neural Networks \cite{schmidhuber2015deep, goodfellow2016deep} have demonstrated extraordinary capability to efficiently model complex non-linear functions in a large variety of applications including HSI classification \cite{santara2017bass, zhu2017deep}. This has motivated us to use Deep Neural Networks with the state-of-the-art BASS-Net architecture of Santara et al. \cite{santara2017bass} as function approximators in all our experiments.\\

Our contributions in this paper can be summarized as follows:
\begin{itemize}
\item We present a PU learning based formulation of the HSI classification problem for material and imaging-platform agnostic large-scale information retrieval. To the best of our knowledge, this is the first work on the investigation of PU Learning for HSI data.\\
\item We design one solution each from the two families of PU learning algorithms -- non-negative risk estimation and PN classification -- and compare their performances on three benchmark HSI datasets.\\
\item We propose a novel spectral-spatial retrieval model for HSI data and use it for intelligent sampling of negative class for PN classification.\\
\item We propose two annotator models that represent the range of labelling patterns of a human annotator and use them to demonstrate the efficacy of our proposed solutions under different spatial distributions of the labelled positive class.\\
\end{itemize}

Section \ref{sec:background} introduces the essential theoretical concepts that we build upon in this paper. Section \ref{sec:framework} gives a detailed description of the proposed framework and approaches to solution. Experimental results are presented in Section \ref{sec:results}. Finally, Section \ref{sec:conclusion} concludes the paper with a summary of our contributions and scope of future work.

\section{Background}
\label{sec:background}
In this section, we present a brief introduction to the essential theoretical concepts used in the rest of the paper.
\subsection{Non-negative Risk Estimation based PU Learning (NNRE-PU)}
Risk estimation based PU learning represents unlabelled data as a weighted combination of P and N data. Following the notation of Kiryo et al. \cite{kiryo2017positive}, let $g:\mathbb{R}^d\rightarrow \mathbb{R}$ be an arbitrary decision function and $l:\mathbb{R}\times \{-1, 1\} \rightarrow \mathbb{R}$ be the loss function such that $l(t,y)$ is the loss incurred on predicting $t$ when the ground truth is $y$. Let $p(\mathbf{X},y)$ denote the joint probability distribution of image-pixels and their labels. The marginal distribution $p(\mathbf{X})$ is where the unlabelled data is sampled from. Let $p_p(\mathbf{X})$ and $p_n(\mathbf{X})$ denote the class-conditionals and $\pi_p$ and $\pi_n$, the prior probabilities of the positive and negative classes respectively. The risk of the decision function $g$ can be written as:
\begin{equation}
\label{Eqn:PN_loss}
    R_{pn}(g) = \pi_p R_p^+(g) + \pi_n R_n^-(g)
\end{equation}

where $R_p^+(g) = \mathbb{E}_{\mathbf{X}\sim p_p}[l(g(\mathbf{X}),+1)]$ and $R_n^-(g) = \mathbb{E}_{\mathbf{X}\sim p_n}[l(g(\mathbf{X}),-1)]$. Rewriting the law of total probability as $\pi_n p_n(\mathbf{X}) = p(\mathbf{X}) - \pi_p p_p(\mathbf{X})$ and substituting in equation \ref{Eqn:PN_loss}, we have the expression for unbiased PU loss:
\begin{equation}
\label{Eqn:unbiased_PU_loss}
    R_{pu}(g) = \pi_p R_p^+(g) - \pi_p R_p^-(g) + R_u^-(g)
\end{equation}

\noindent where $R_p^-(g) = \mathbb{E}_{\mathbf{X}\sim p_p}[l(g(\mathbf{X}),-1)]$ and $R_u^-(g) = \mathbb{E}_{\mathbf{X}\sim p}[l(g(\mathbf{X}),-1)]$. In unbiased PU learning \cite{elkan2008learning, du2014analysis, du2015convex}, the goal is to minimize an empirical estimate of this risk $R_{pu}(g)$ (with the expectations replaced by sample averages) to find the optimal decision function $g$. Unfortunately, the empirical estimators of risk used in unbiased PU learning have no lower bound although the original risk objective in equation \ref{Eqn:PN_loss} is non-negative \cite{kiryo2017positive}. Minimization of the empirical risk tends to drive the objective negative without modeling anything meaningful especially when high capacity function approximators like deep neural networks are used to model $g$. Kiryo et al. \cite{kiryo2017positive} propose the following biased, yet optimal, non-negative risk estimator to address this problem:
\begin{equation}
\label{Eqn:nnPU_loss}
    \hat{R}_{nn-PU}(g) = \pi_p \hat{R}_p^+(g) + \max\left(0, \hat{R}_u^-(g) - \pi_p \hat{R}_p^-(g)\right)
\end{equation}
where $\hat{R}$ denotes an empirical estimate of actual risk, $R$. For the ease of training of a neural network classifier, with no loss in theoretical correctness, we represent the negative class by $0$ instead of $-1$ in our experiments.\\

\subsection{Non-Local Total Variation}
\label{sec:PDHG}
Non-Local Total Variation (NLTV) is an unsupervised clustering objective demonstrated on hyperspectral data by Zhu et al. \cite{zhu2017unsupervised}. Following the notation used by the authors, let $\Omega \subset \mathcal{I}$ be a region in a hyperspectral scene. Let $L^2(\Omega)$ be a Hilbert space. let $u:\Omega \rightarrow \mathbb{R},\; u\in L^2(\Omega)$, be the labelling function of a cluster such that the larger the value of $u(\mathbf{X})$, the higher is the likelihood of a pixel $\mathbf{X}$ belonging to that cluster. Let $d:\mathcal{I}\rightarrow\mathcal{I}$ be a measure of divergence between two given pixels such that a lower value of $d$ implies more resemblance. \emph{Non-local derivative} is defined as:
\begin{equation}
\label{Eqn:nonlocal_derivative}
    \frac{\partial u}{\partial y}(x) := \frac{u(y)-u(x)}{d(x,y)}, \forall x,y \in \Omega
\end{equation}
\emph{Non-local weight} is defined as $w(x,y)=d^{-2}(x,y)$. The expression for non-local derivative in equation \ref{Eqn:nonlocal_derivative} can be rewritten in terms of non-local weight as:
\begin{equation}
\label{Eqn:nonlocal_derivative_2}
    \nabla_w u(x)(y) = \frac{\partial u}{\partial y}(x) := \sqrt{w(x,y)}(u(y)-u(x))
\end{equation}
The Non-Local Total Variation (NLTV) objective is given by:
\begin{equation}
\label{Eqn:NLTV_obj}
    \min_u E(u) = ||\nabla_w u||_{L^1} + \lambda S(u)
\end{equation}
$S(u)$ is a \emph{data fidelity term} representing the clustering objective and $||\nabla_w u||_{L^1}$ is the Total Variation regularizer. The parameter $\lambda$ controls the amount of regularization. The authors of \cite{zhu2017unsupervised} present a linear and a quadratic model of this objective, depending upon the design of $S(u)$. They also apply the Primal Dual Hybrid Gradient (PDHG) algorithm \cite{chambolle2011first} for minimization of these objectives and show encouraging results on hyperspectral image data. We use the quadratic model of the NLTV objective in our experiments.

\vspace{0.5cm}
\subsection{BASS-Net architecture}
\begin{figure*}
\centering
	\includegraphics[width=0.8\textwidth]{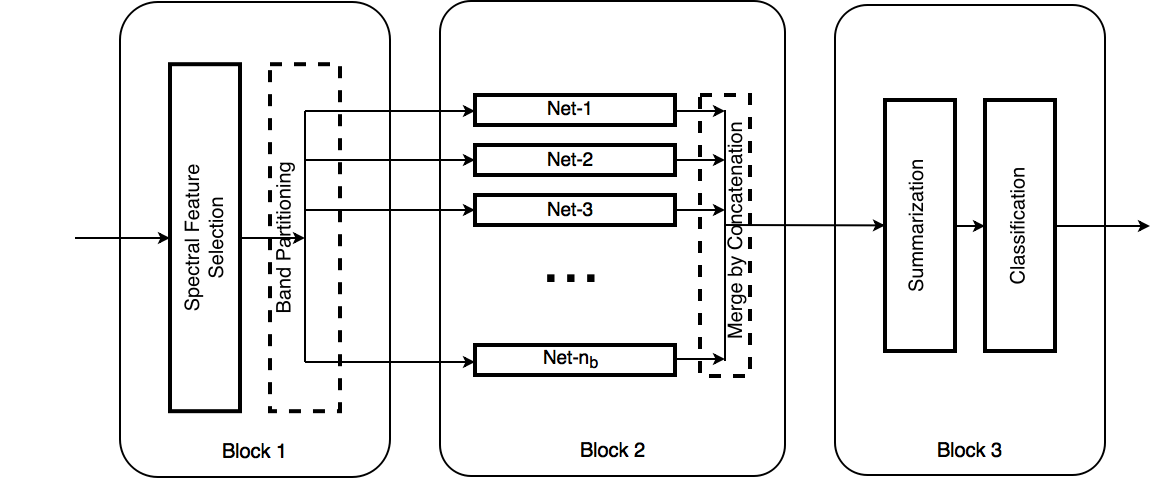}
	\caption{Schematic diagram of the BASS-Net architecture \cite{santara2017bass}.}
	\label{fig:BASS-Net}
\end{figure*}

Band-Adaptive Spectral-Spatial feature learning Network (BASS-Net) \cite{santara2017bass} is a deep neural network architecture for end-to-end supervised classification of Hyperspectral Images. Hyperspectral image classification poses two unique challenges: a) curse of dimensionality resulting from large number of spectral dimensions and scarcity of labelled training samples, and, b) large spatial variability of spectral signature of materials. The BASS-Net architecture is extremely data efficient thanks to extensive parameter sharing along the spectral dimension and is capable of learning highly non-linear functions from a relatively small number of labelled training examples. Also, it uses spatial context to account for spatial variability of spectral signatures. BASS-Net shows state-of-the-art supervised classification accuracy on benchmark HSI datasets like Indian Pines, Salinas and University of Pavia. Figure \ref{fig:BASS-Net} shows a schematic diagram of the BASS-Net architecture.

\section{PU Classification of Hyperspectral Images}
\label{sec:framework}
\begin{algorithm}[t]
\begin{algorithmic}[1]
\caption{NNRE-PU based Information Retrieval in a HSI scene}
\label{algo:NNRE-PU}
\INPUT \ Hyperspectral scene $\mathcal{I}$, labelled positive set $\mathcal{I}_{L+}$, prior probability of positive class $\pi_p$.
\OUTPUT Map of all predicted occurrences of the positive class in the whole scene\\
Sample unlabelled data uniformly from $\mathcal{I}_U$.\\
Train a BASS-Net classifier to minimize the NNRE-PU objective defined in Eq. \ref{Eqn:nnPU_loss}.\\
Classify every pixel of $\mathcal{I}$ using the trained neural network to generate the output map.
\end{algorithmic}

\end{algorithm}

\begin{algorithm}[!h]
\begin{algorithmic}[1]
\caption{PN-PU based Information Retrieval in a HSI scene}
\label{algo:NN-PU}
\INPUT \ Hyperspectral scene $\mathcal{I}$, labelled positive set $\mathcal{I}_{L+}$, \ number of clusters $\mathcal{C}$ for clustering using PDHG algorithm.
\OUTPUT \ Map of all predicted occurrences of the positive class in the whole scene.\\
Run PDHG for $\mathcal{C}$ clusters\footnotemark.\\
Sample the same number of unlabelled negative pixels as labelled positive pixels from the distribution $Pr_{-}(\mathbf{X}|\mathbf{X}\in \mathcal{I}_U) = 1 - Pr_{+}(\mathbf{X}|\mathbf{X}\in \mathcal{I}_U)$ where $Pr_{+}(\mathbf{X}|\mathbf{X}\in \mathcal{I}_U)$ is given by our spectral-spatial retrieval model defined in equation \ref{Eqn:hybrid-spectral-spatial}.\\
Train a BASS-Net classifier to minimize cross-entropy loss.\\
Classify each pixel of $\mathcal{I}$ using the trained BASS-Net to generate the output map.
\end{algorithmic}
\end{algorithm}
\footnotetext{If the output of clustering is available from a previous run of this algorithm on the same HSI scene, we reuse the results.}



In this section we describe the proposed PU learning algorithms for HSI classification. As mentioned in the previous section, we use the BASS-Net architecture of Santara et al. \cite{santara2017bass} as function approximators, whenever required, in our pipeline. The input to the network is a pixel $\mathbf{X}$ from the image $\mathcal{I}$ with its $p\times p$ neighborhood (for spatial context) in the form of a $p \times p \times N_c$ volume, where $N_c$ is the number of channels in the input image. The output is the predicted class label $\hat{y} \in \{-1,1\}$ for $\mathbf{X}$. The specific configuration of BASS-Net that we use in our experiments is Configuration 4 of Table 1 of Santara et al. \cite{santara2017bass}. As BASS-Net was originally made for multi-class classification, it used a softmax layer at the output and optimized a multi-class categorical cross entropy loss defined as follows.
\begin{equation}
\label{Eqn:cross-entropy-loss}
    \mathcal{L}_{\times-entropy}=\sum_{c=1}^M y_{o, c} \log p_{o, c}
\end{equation}
Where $M$ is the total number of classes, $y_{o, c}$ is a binary indicator function which returns $1$ when class $c$ is the correct classification of observation $o$ and $0$ otherwise, and $p_{o, c}$ denotes the predicted probability of observation $o$ belonging to class $c$. In PU learning, we work with binary classifiers. Hence we replace the softmax layer with a sigmoid layer and use the binary cross entropy loss function (equation \ref{Eqn:cross-entropy-loss} with $M=2$) for training.\\ 

Let $\mathcal{I}_{L+}$ denote the set of labeled positive data points and $\mathcal{I}_{U}$ be the unlabeled data points such that $\mathcal{I}=\mathcal{I}_{L+}\cup \mathcal{I}_U; \; \mathcal{I}_{L+}\cap \mathcal{I}_U=\phi$. Let $\pi_p$ denote the prior probability of the positive class in the entire image. \\

In the first set of experiments, we implement the NNRE-PU learning algorithm of Kiryo et al. \cite{kiryo2017positive} (Algorithm \ref{algo:NNRE-PU}). As the true value of $\pi_p$ is unknown for an arbitrary HSI scene, the user (who is expected to have some domain knowledge) has to make an estimate of $\pi_p$ from visual inspection of the image. We study how the performance of the classifier varies with perturbations to the true value of $\pi_p$.\\

In our second set of experiments, we evaluate PN classification based PU learning (PN-PU). Algorithm \ref{algo:NN-PU} describes the workflow. A novel spectral-spatial retrieval model, described in Section \ref{sec:hybrid-spectral-spatial}, is used to model the conditional positive class probability $Pr_{+}(\mathbf{X}|\mathbf{X}\in \mathcal{I}_U)$ of an unlabelled pixel $\mathbf{X}$. Negative samples are drawn from $Pr_{-}(\mathbf{X}|\mathbf{X}\in \mathcal{I}_U) = 1-Pr_{+}(\mathbf{X}|\mathbf{X}\in \mathcal{I}_U)$. A PN classifier having the BASS-Net architecture is then trained on the labelled positive and sampled negative examples.

\subsection{Heuristic-based Probability Estimates}
\label{sec:heuristics}
We explore two heuristics for modeling $Pr_{+}(\mathbf{X}|\mathbf{X}\in \mathcal{I}_U)$, the conditional positive class probability of an unlabelled pixel $X$.

    \subsubsection{Spatial distance based} According to this heuristic, the conditional probability of an unlabelled pixel belonging to the positive class decreases with its Euclidean distance from the nearest labelled positive sample. Let $d_+(\mathbf{X}) = min_{\mathbf{X}^'\in \mathcal{I}_{L+}} ||\mathbf{X}-\mathbf{X}^'||_2$ be the Euclidean distance of $\mathbf{X}$ from the nearest labelled positive pixel. Then we have,
    \begin{equation}
    \label{eqn:spatial_heuristic}
        Pr_{+}(\mathbf{X}|\mathbf{X}\in \mathcal{I}_U) = \frac{1}{1+\exp(\frac{d_+(\mathbf{X})-b}{T})}
    \end{equation}
    where baseline $b$ and temperature $T$ are hyperparameters.
    This heuristic draws from the intuition of spatial continuity of a material.\\ 
    The primary drawbacks of this heuristic are a) it assumes that the user labels positive pixels uniformly over all occurrences of the positive class in the scene, and b) it does not use any notion of spectral similarity of pixels. Imagine a scene in which the positive class occurs in several disconnected locations of the scene, far away from one another. If the user only labels pixels in one of these locations, the positive pixels from the other locations would have a high chance of being wrongly sampled as negative class by this heuristic -- thus affecting the sensitivity of the classifier.
    
    \subsubsection{Spectral similarity based} We use an unsupervised segmentation algorithm (PDHG \cite{zhu2017unsupervised}, in our experiments), to cluster the hyperspectral scene into a set of $C$ clusters based on spectral similarity. Suppose an unlabelled sample belongs to a cluster of size $n$. If $m$ of the samples from its cluster were labelled positive by the user, then, the probability of the unlabelled pixel belonging to the positive class is given by:
    \begin{equation}
    \label{eqn:spectral_heuristic}
        Pr_{-}(\mathbf{X}|\mathbf{X}\in \mathcal{I}_U) = \frac{m+\epsilon}{n}
    \end{equation}
    Where, $\epsilon$ is a small positive number ($10^{-4}$ in our experiments).
    This way, we sample more pixels of the negative class from regions of the scene that differ significantly in spectral characteristics from the labelled positive class.
    The main drawback of this heuristic stems from its assumption that each cluster containing positive pixels is likely to contain some user-labelled pixels. This is possible only under two conditions: a) the spectral uniformity of the positive class is high enough for the unsupervised segmentation algorithm to include all the positive pixels in the same cluster, or, b) the user labels the positive pixels uniformly over all the different regions of the HSI scene in which the positive class occurs. Additionally, under this model, every pixel in a cluster gets assigned the same probability of being sampled for the negative class regardless of its spatial-proximity to the labelled positive samples. This can directly affect the specificity of the classifier. 
    
\subsection{Spectral-spatial Retrieval Model}
\label{sec:hybrid-spectral-spatial}
The spectral and spatial heuristics make certain assumptions about material distribution and the behavior of the annotator. Although these assumptions seldom hold completely, they do hold true to a certain degree in natural HSI scenes. Our goal is to design a retrieval model that outputs a lower-bound on the estimate of $Pr_{+}(\mathbf{X}|\mathbf{X}\in \mathcal{I}_U)$ whenever one or more of these assumptions are violated in an image. A multiplicative combination of the spatial (equation \ref{eqn:spatial_heuristic}) and spectral (equation \ref{eqn:spectral_heuristic}) factors described in Section \ref{sec:heuristics} achieves this goal and compensates for the drawbacks of the individual heuristics. The conditional probability of an unlabelled pixel $X$ belonging to the positive class under this retrieval model is given by:
\begin{equation}
\label{Eqn:hybrid-spectral-spatial}
    Pr_{+}(\mathbf{X}|\mathbf{X}\in \mathcal{I}_U) = \left(\frac{m+\epsilon}{n}\right)\times
    \left(\frac{1}{1+\exp(\frac{d_+(\mathbf{X})-b}{T})}\right)
\end{equation}

\noindent We evaluate our retrieval model on three benchmark HSI datasets and two annotation models that simulate the behavior of a human annotator.


\section{Experimental Results}
\label{sec:results}
In this section we compare the performances of the methods proposed in Section \ref{sec:framework}.

\subsection{Data Sets}
\label{sec:data sets}
We perform our experiments on three popular hyperspectral image classification data sets -- Indian Pines \cite{PURR1947}, Salinas, and Pavia University scene\footnote{http://www.ehu.eus/ccwintco/index.php?title=Hyperspectral\_Remote\_ Sensing\_Scenes}. Some classes in the Indian Pines data set have very few samples. We reject those classes and select the top $9$ classes by population for experimentation. The problem of insufficient samples is less severe for Salinas and U. Pavia and all the classes are taken into account. We choose \emph{Corn-notill} (class $2$), \emph{Stubble} (class $6$) and \emph{Asphalt} (class $1$) as positive classes for Indian Pines, Salinas and U. Pavia datasets respectively. These materials appear in multiple disconnected patches in their corresponding scenes. This puts to test the ability of our methods to model the dramatic spatial variability of spectral signature in HSI scenes. Additionally, these materials form mid-sized classes in their corresponding datasets. Hence the results we obtain about them from our experiments are statistically significant. We sample $10\%$ of the pixels of the positive class using one of the annotation models described in Section \ref{sec:annotation_models} to construct the labelled positive set, $\mathcal{I}_{L+}$. The rest of the pixels constitute the set of unlabelled samples, $\mathcal{I}_{U}$. For NNRE-PU experiments we uniform-randomly sample $5000$ points from $\mathcal{I}_{U}$ for use in training. In the PN classification experiments, an equal number of negative samples as the labelled positive set are drawn from $Pr_{-}(\mathbf{X}|\mathbf{X}\in \mathcal{I}_U) = 1 - Pr_{+}(\mathbf{X}|\mathbf{X}\in \mathcal{I}_U)$ where $Pr_{+}(\mathbf{X}|\mathbf{X}\in \mathcal{I}_U)$ is given by equation \ref{Eqn:hybrid-spectral-spatial}. As different frequency channels have different dynamic ranges, their values are normalized to the range $[0,1]$ using the transformation $f(\cdot)$ defined in equation \ref{Eqn:inp_norm}, where $x$ denotes the random variable corresponding to the pixel values of a given channel.
\begin{equation}
\label{Eqn:inp_norm}
f(x) = \frac{x - min(x)}{max(x)-min(x)}
\end{equation}

\subsection{Annotation Models}
\label{sec:annotation_models}
We explore two annotation models for constructing the labelled positive set, $\mathcal{I}_{L+}$: a) \emph{uniform}, and b) \emph{blob}. Imagine that the positive class forms $N$ connected components in an HSI scene -- where two pixels are considered connected if and only if they are adjacent to each other. The uniform annotation model samples $\mathcal{I}_{L+}$ uniformly from all the connected components. It models the case in which the user labels positive samples uniformly across all instances of spatial occurrence of the positive class. The blob annotation model, on the other hand, models the more practical case in which the user labels a small blob of positive pixels in one location of the image. We implement blob annotation model by starting at a random positive sample, adding it to $\mathcal{I}_{L+}$ and expanding $\mathcal{I}_{L+}$ by searching and adding the adjoining positive pixels in a breadth first fashion. The sampler never leaves a connected component of positive pixels until all the pixels have been included in $\mathcal{I}_{L+}$. After that, it shifts to a random positive pixel in a different connected component and repeats the process until the requisite number of positive samples have been drawn.

\begin{table}
\caption{data sets used}
\label{table:data sets}
\centering
\begin{tabular}{l|p{1.7cm}|p{1.5cm}|p{1.4cm}}

     & Indian Pines & Salinas & U. Pavia  \\ \hline\hline 
Sensor & AVIRIS     & AVIRIS  & ROSIS \\ \hline 
Place  & Northwestern Indiana & Salinas Valley California & Pavia, Northern Italy \\ \hline 
Frequency Band & $0.4$-$0.45 \mu m$ & $0.4$-$0.45 \mu m$ & $0.43$-$0.86 \mu m$ \\ \hline 
Spatial Resolution & $20m$ & $20m$ & $1.3m$ \\ \hline 
No. of Channels & 220 & 224 & 103 \\ \hline 
No. of Classes & 16 & 16 & 9 \\ \hline
\end{tabular}
\end{table}

\subsection{Evaluation Metrics}
We evaluate our PU learning algorithms in terms of precision, recall, F-score, and area under the receiver operating characteristics curve (AUC) \cite{powers2011evaluation}. 

\subsection{Operating Point and Hyperparameter Selection}
We choose the operating point of our classifiers to minimize the expected cost of mis-classification of a point in the Receiver Operating Curve (ROC) space \cite{ROC_article} given by:
\begin{equation}
    \mathcal{C} = (1-\pi_p)\alpha x + \pi_p \beta (1-y)
\end{equation}
$x$ and $y$ are coordinates of the ROC space, and $\alpha$ and $\beta$ are the costs of a false positive and false negative respectively. We assume $\alpha=\beta$. The expectation is performed on a validation set consisting of $7\%$ of the unlabelled samples in the dataset. The solution for our operating point is the point on the ROC curve that lies on a line of slope $\frac{1-\pi_p}{\pi_p}$ closest to the north-west corner, $(0, 1)$, of the ROC plot. The temperature $T$ and baseline $b$ hyperparameters are also tuned on the validation set through a grid search and presented in Table \ref{table:temp_baseline}. The number of clusters $\mathcal{C}$ for PDHG algorithm is set to the number of classes in the respective datasets given in Table \ref{table:data sets}. All neural networks are trained for $100$ epochs or till over-fitting sets in (validation loss starts increasing), whichever happens first.

\begin{table}[]
\caption{Temperature and Baseline settings}
\label{table:temp_baseline}
\centering
\begin{tabular}{l|c|c|c|}
\cline{2-4}
                                                 & Indian Pines & Salinas & U. Pavia \\ \hline
\multicolumn{1}{|l|}{Temperature ($T$)} & $24$                  & $22$             & $14$              \\ \hline
\multicolumn{1}{|l|}{Baseline ($b$)}    & $32$                  & $26$             & $26$              \\ \hline
\end{tabular}
\end{table}

\subsection{Implementation Platform}
The algorithms have been implemented in Python using the Chainer deep learning library \cite{tokui2015chainer} for highly optimized training of neural networks. As every execution of the proposed algorithms involve training of a neural network, the computational demand is high. Chainer has native support for multi-core CPU and multi-GPU parallelism. This makes it a natural choice for our application. As Chainer only supports input images with even number of channels, we append a new channel with all zeros to the Pavia University HSI scene. Our code is available open-source on GitHub\footnote{https://github.com/HSISeg}.

\begin{table}[]
\caption{PU classification results}
\label{table:results}
\centering
\begin{tabular}{|l|l|c|c|c|c|}
\hline
\multicolumn{1}{|c|}{\multirow{3}{*}{Dataset}} & \multicolumn{1}{c|}{\multirow{3}{*}{Metric}} & \multicolumn{4}{c|}{Retrieval Model} \\ \cline{3-6} 
\multicolumn{1}{|c|}{} & \multicolumn{1}{c|}{} & \multicolumn{2}{c|}{Uniform Sampling} & \multicolumn{2}{c|}{Blob Sampling} \\ \cline{3-6} 
\multicolumn{1}{|c|}{} & \multicolumn{1}{c|}{} & NNRE-PU & PN & NNRE-PU & PN \\ \hline
\multirow{4}{*}{Indian Pines} & AUC & $0.90$ & $0.85$ & $0.83$ & $0.67$ \\ \cline{2-6} 
 & Precision & $83.0\%$ & $48\%$ & $46\%$ & $48\%$ \\ \cline{2-6} 
 & Recall & $77.1\%$ & $83.1\%$ & $81.1\%$ & $42\%$ \\ \cline{2-6} 
 & F-score & $0.80$ & $0.61$ & $0.59$ & $0.45$ \\ \hline
\multirow{4}{*}{Salinas} & AUC & $0.98$ & $0.99$ & $0.99$ & $0.91$ \\ \cline{2-6} 
 & Precision & $99.8\%$ & $99.7\%$ & $99.9\%$ & $99.9\%$ \\ \cline{2-6} 
 & Recall & $98.3\%$ & $99.9\%$ & $97.7\%$ & $82\%$ \\ \cline{2-6} 
 & F-score & $0.99$ & $0.99$ & $0.98$ & $0.89$ \\ \hline
\multirow{4}{*}{Pavia U} & AUC & $0.90$ & $0.95$ & $0.91$ & $0.93$ \\ \cline{2-6} 
 & Precision & $91\%$ & $96.4\%$ & $80.2\%$ & $60\%$ \\ \cline{2-6} 
 & Recall & $88.6\%$ & $76\%$ & $86\%$ & $91\%$ \\ \cline{2-6} 
 & F-score & $0.89$ & $0.85$ & $0.83$ & $0.72$ \\ \hline
\end{tabular}
\end{table}

\begin{figure}
    \centering
    \subfloat[]{\includegraphics[width=0.9\linewidth]{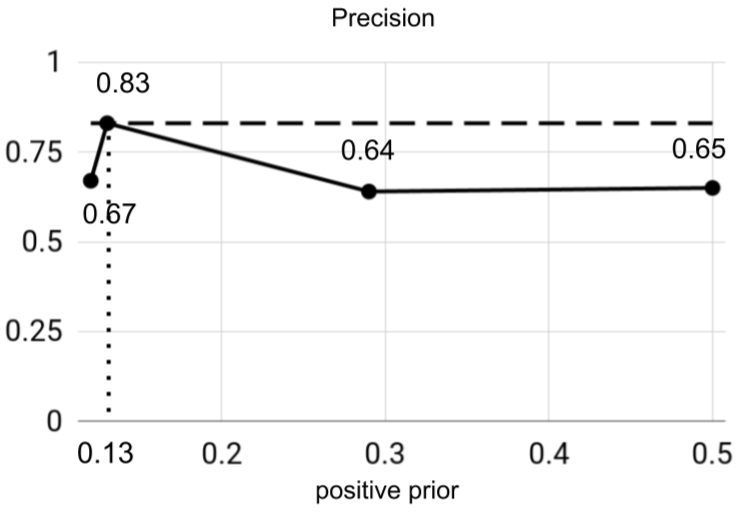}}\\
    \subfloat[]{\includegraphics[width=0.9\linewidth]{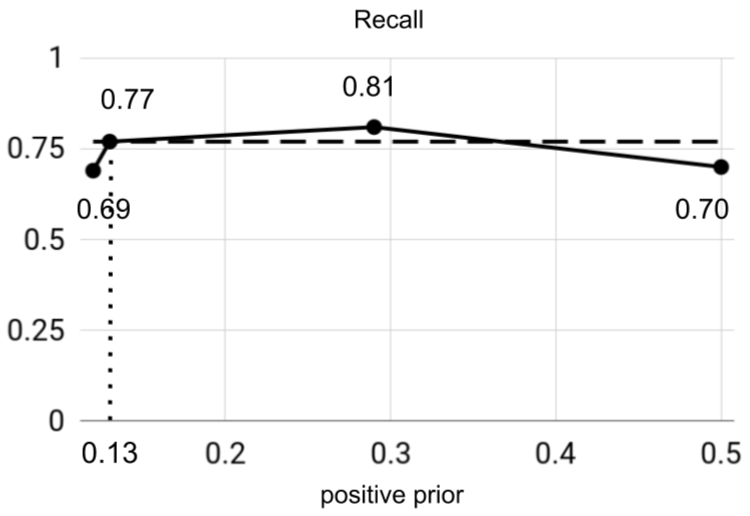}}
	\caption{Variation of test (a) precision and (b) recall of NNRE-PU on the Indian Pines data-set with input value of positive prior $\pi_p$. The dashed line denotes performance for the true value of $\pi_p = 0.13$.}
	\label{fig:precision_recall_vs_pi_p_PU}
\end{figure}%

\begin{table*}[!h]
\caption{Sample outputs}
\label{table:sample_outputs}
\centering
\begin{tabular}[t]{p{0.15\textwidth}P{0.25\textwidth}P{0.25\textwidth}P{0.25\textwidth}}
\textbf{Experiment Name}                                      & \textbf{Training Data}                                                              & \textbf{Prediction on the Test Set}                                                          & \textbf{Test Set Confusion Map}                                                                    \\ 
                                                              & \includegraphics[width=0.6\linewidth]{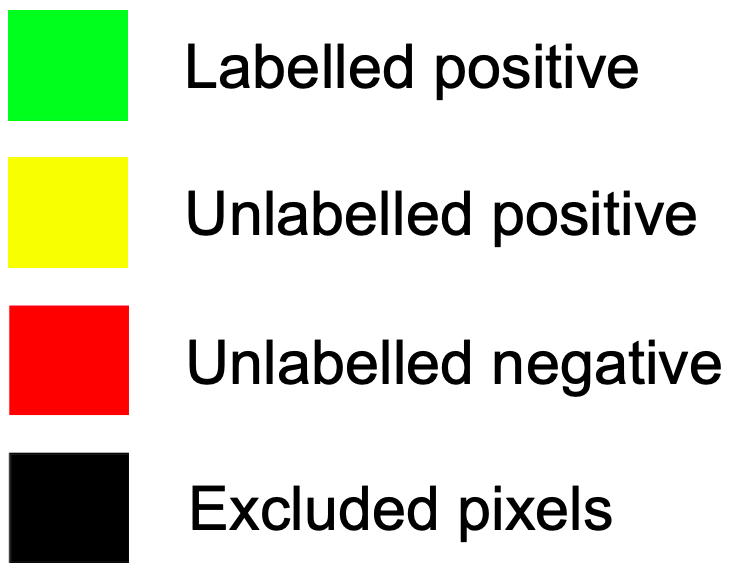} & \includegraphics[width=0.6\linewidth]{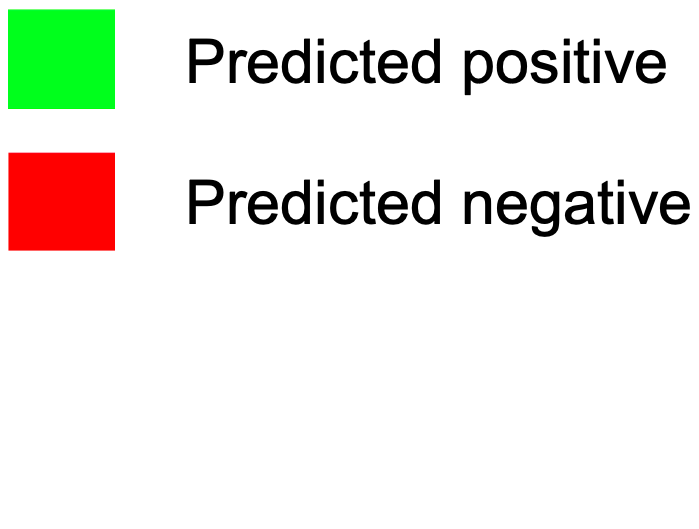} & \includegraphics[width=0.55\linewidth]{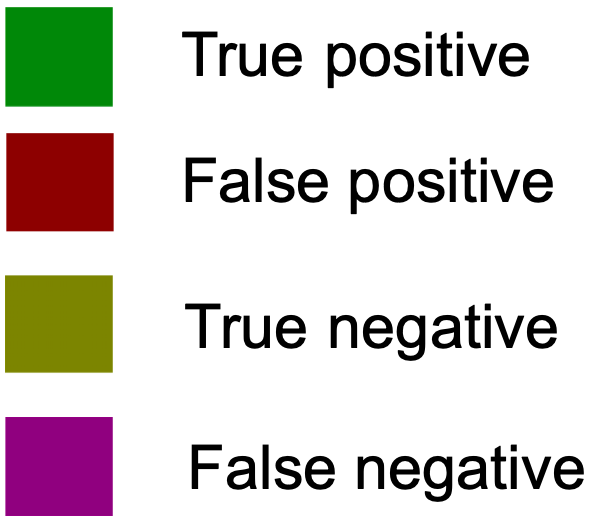}\\ 
\vspace{-2cm}NNRE-PU on Indian Pines with uniform sampling retrieval model & \includegraphics[width=0.9\linewidth, height=0.9\linewidth]{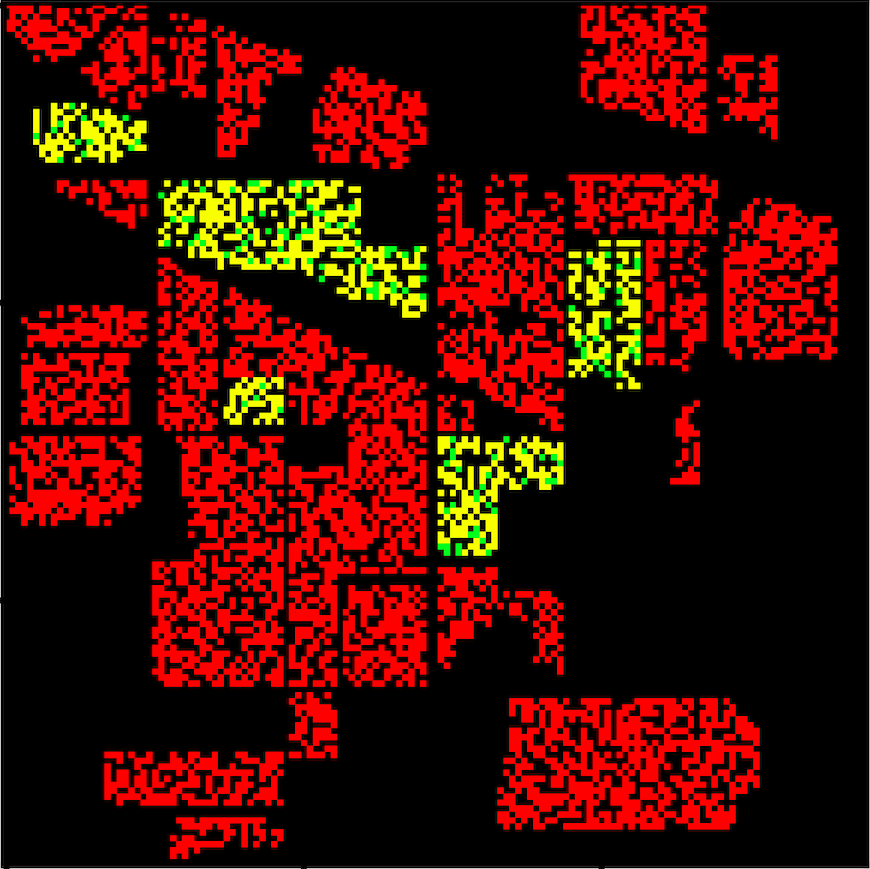} & \includegraphics[width=0.9\linewidth, height=0.9\linewidth]{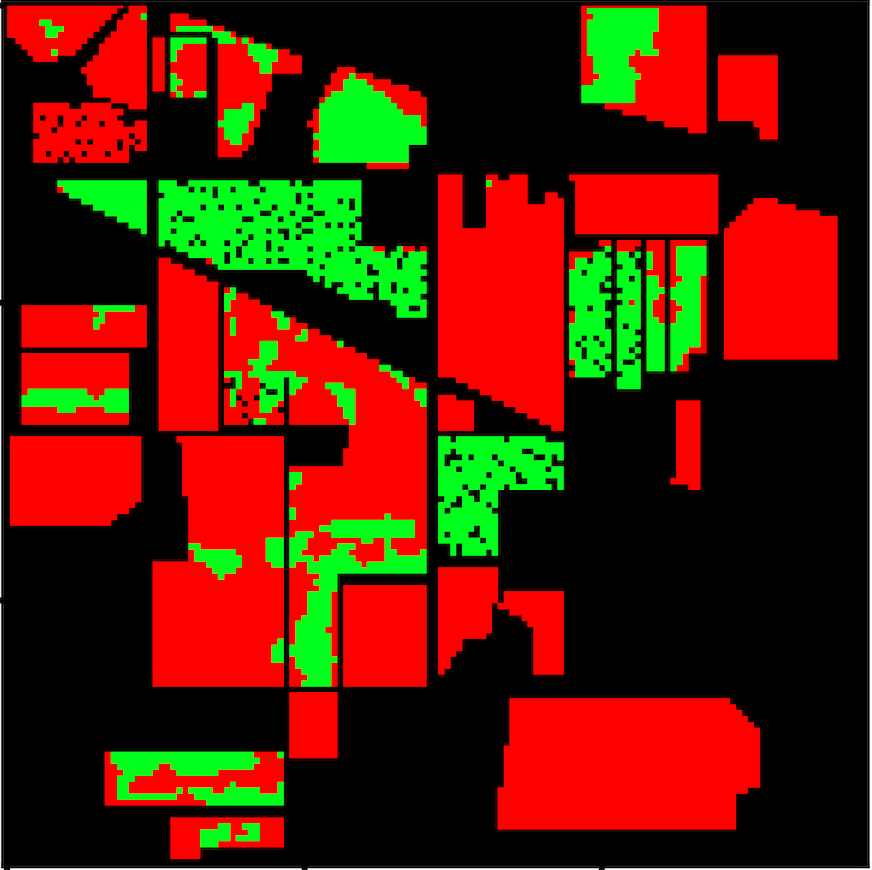} & \includegraphics[width=0.9\linewidth, height=0.9\linewidth]{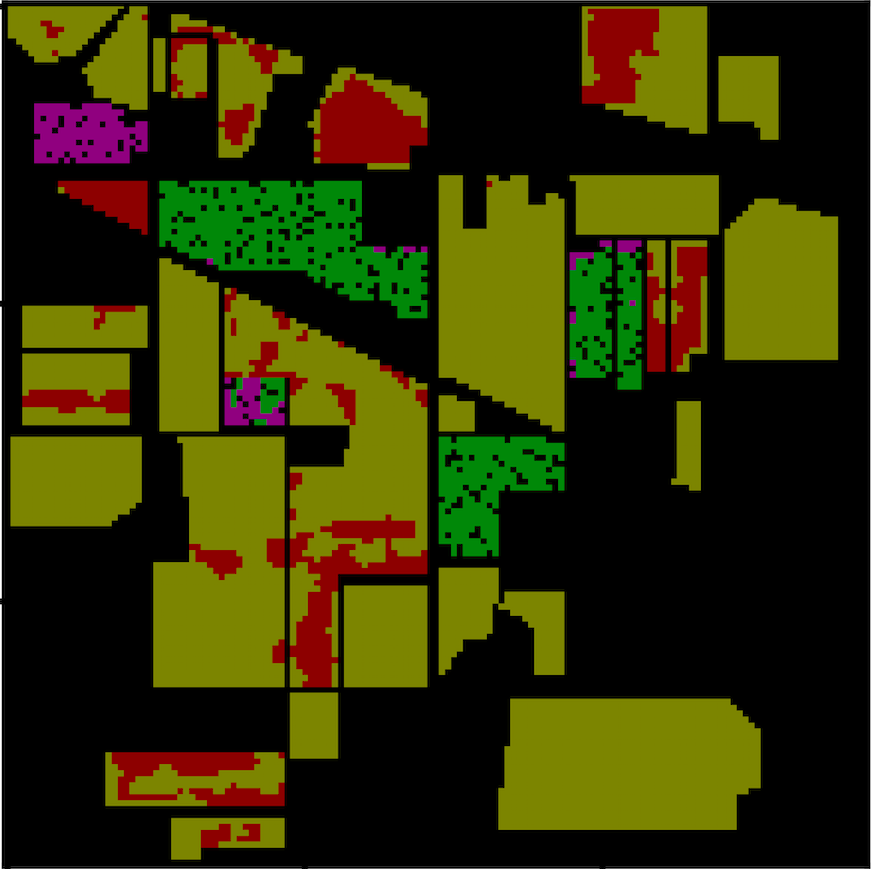} \\ 
\vspace{-2cm}PN-PU on Indian Pines with uniform sampling retrieval model   & \includegraphics[width=0.9\linewidth, height=0.9\linewidth]{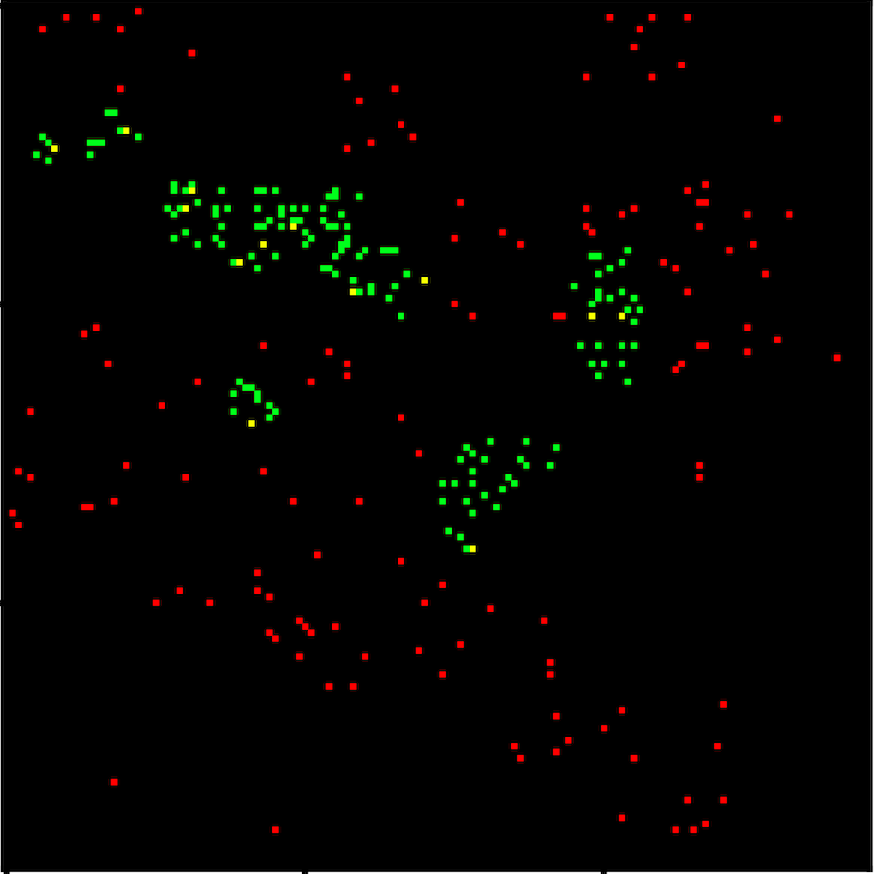}   & \includegraphics[width=0.9\linewidth, height=0.9\linewidth]{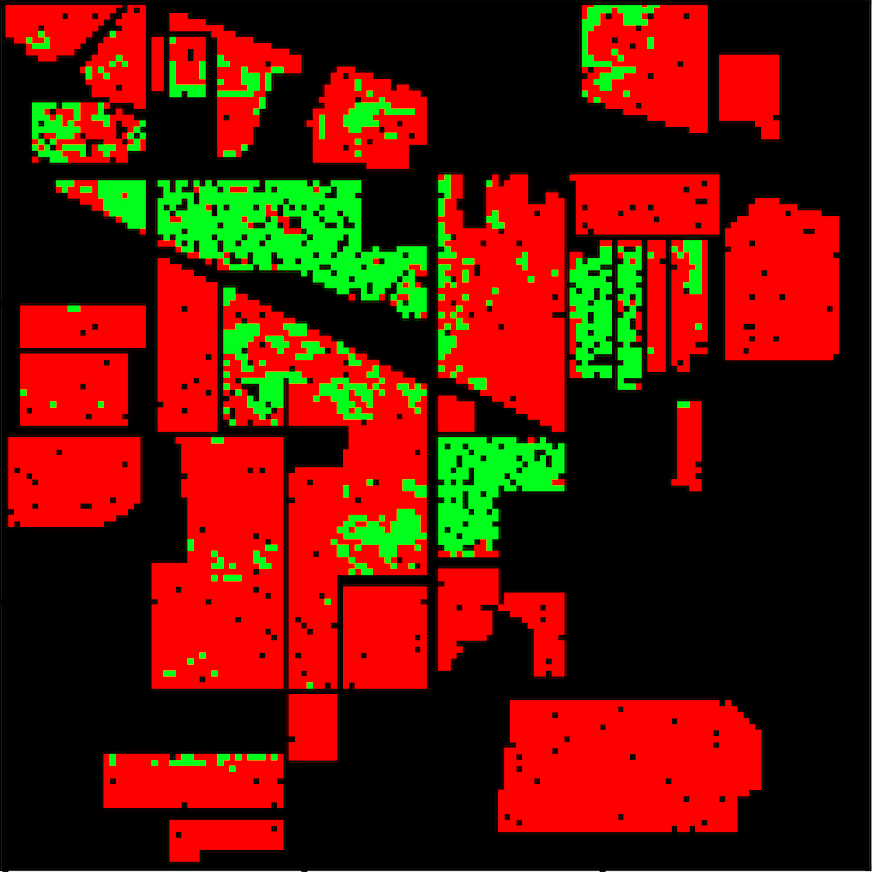}   & \includegraphics[width=0.9\linewidth, height=0.9\linewidth]{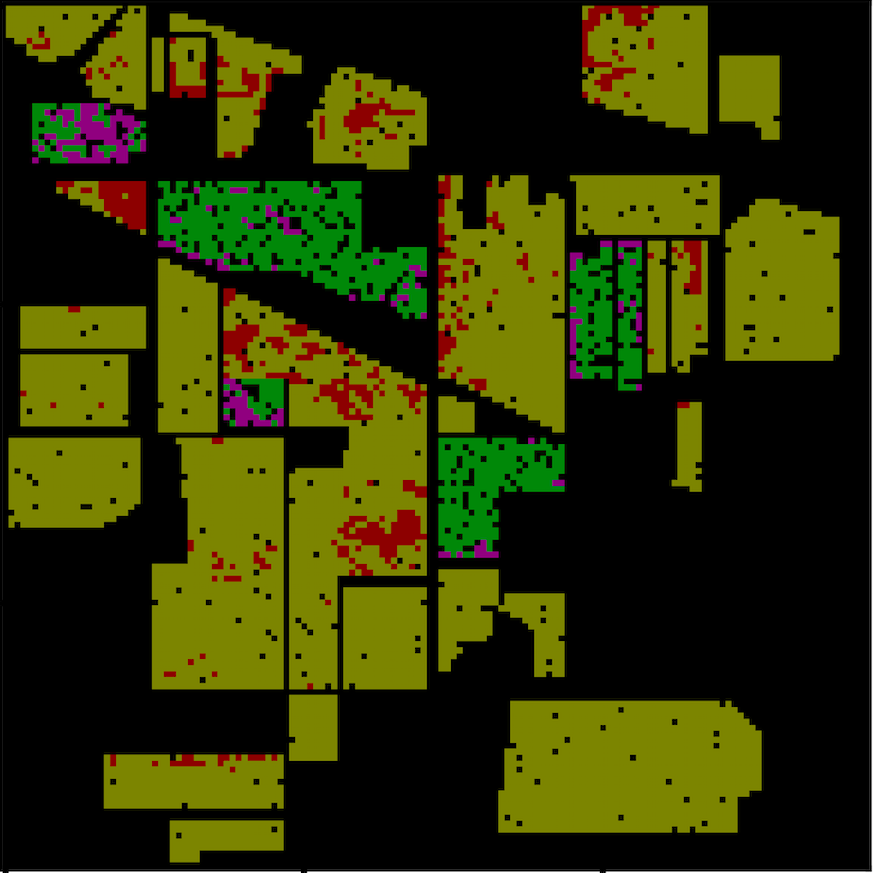}   \\ 
\vspace{-2cm}NNRE-PU on Salinas with blob sampling retrieval model.         & \includegraphics[width=0.9\linewidth, height=0.9\linewidth]{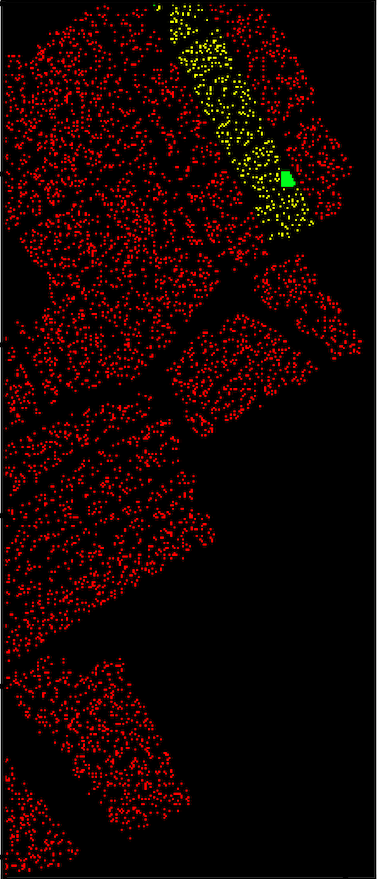}        & \includegraphics[width=0.9\linewidth, height=0.9\linewidth]{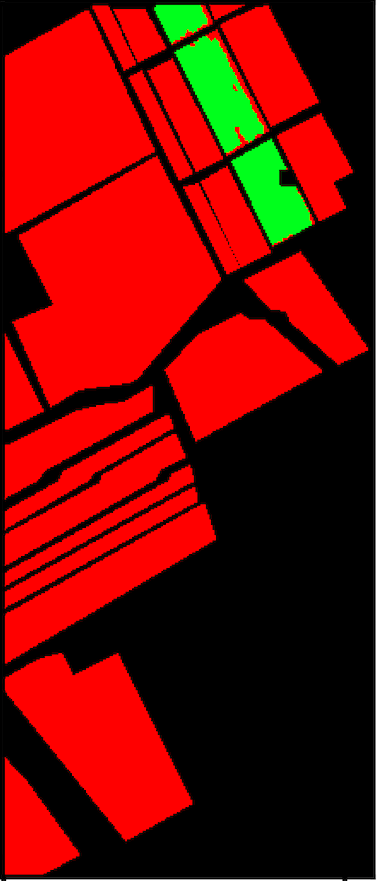}          & \includegraphics[width=0.9\linewidth, height=0.9\linewidth]{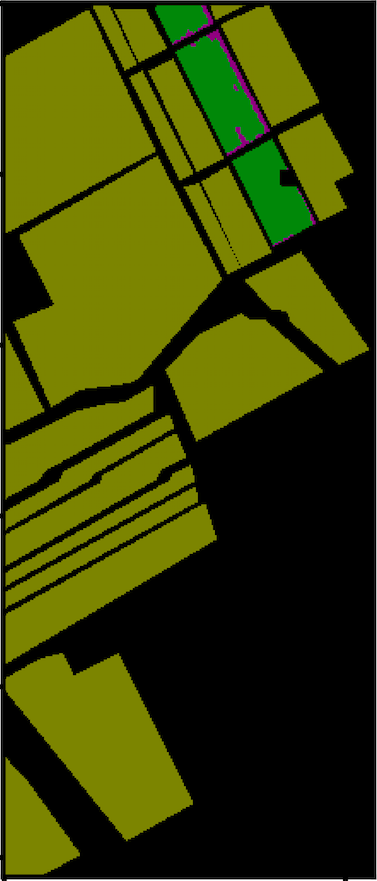}          \\ 
\vspace{-2cm}PN-PU on Pavia University with blob sampling retrieval model.  & \includegraphics[width=0.9\linewidth, height=0.9\linewidth]{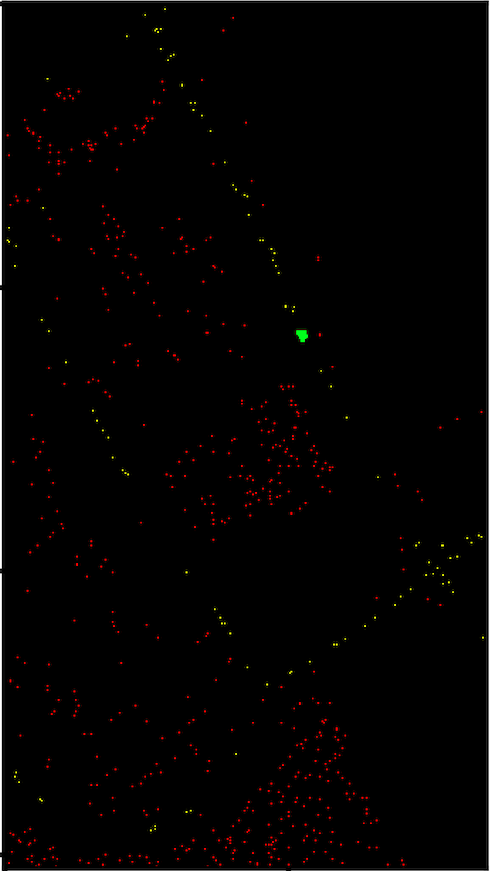}           & \includegraphics[width=0.9\linewidth, height=0.9\linewidth]{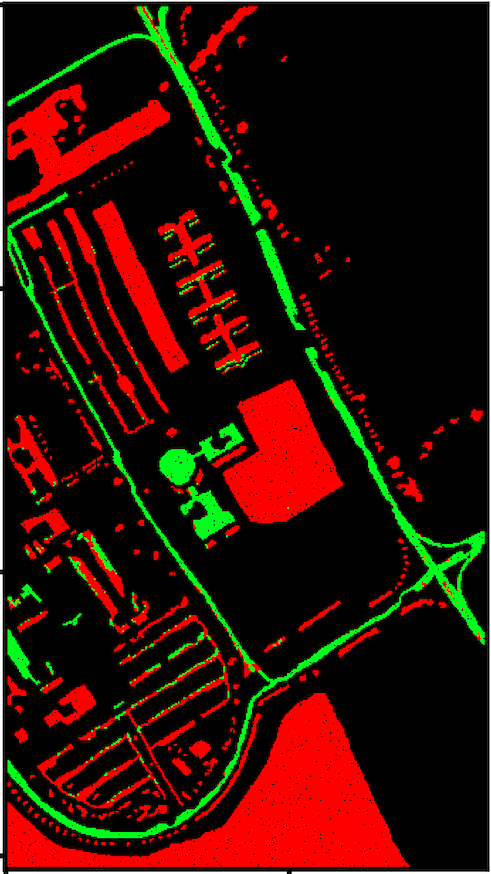}           & \includegraphics[width=0.9\linewidth, height=0.9\linewidth]{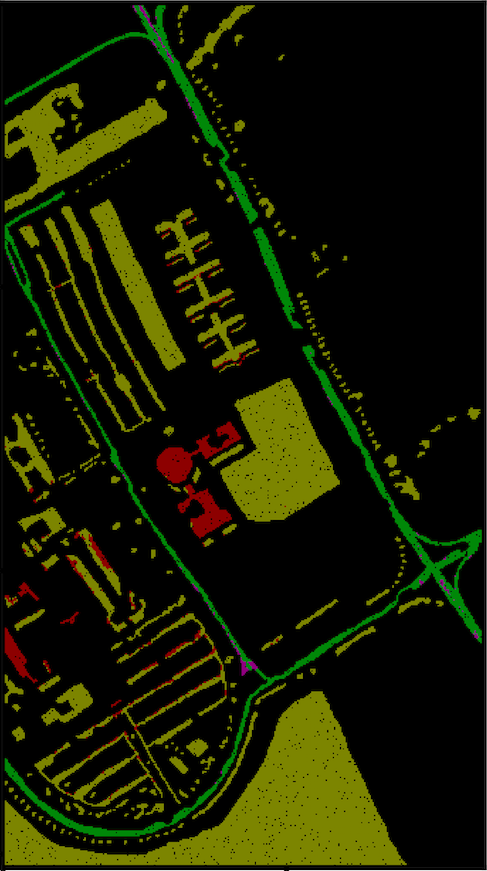}           \\ 
& & & \\
& & & \\
& & & \\
& & & \\
\end{tabular}
\end{table*}
\subsection{Results and Discussion}
\label{sec:discussion}
Table \ref{table:results} presents the numerical outcomes of our experiments. The NNRE-PU numbers correspond to the true values of $\pi_p$ for each dataset. Table \ref{table:sample_outputs} shows the visual outputs of some sample runs of the proposed algorithms. We make the following observations:
\begin{itemize}
    \item The performance of NNRE-PU is highly sensitive to the value of $\pi_p$ supplied by the user. Figure \ref{fig:precision_recall_vs_pi_p_PU} shows the variation of precision and recall with the supplied value of $\pi_p$ for Indian Pines dataset with uniform sampling of the positive class. This is a major drawback of the NNRE-PU approach because it is hard -- even for an expert -- to provide an accurate estimate $\pi_p$ for the query material in an arbitrary HSI scene.\\
    
    \item PN-PU, with the right hyperparameter settings in the spectral-spatial retrieval model, gives performance comparable with NNRE-PU -- although there is no clear trend of supremacy of any one method across all the different HSI datasets and annotation models. Unlike NNRE-PU, PN-PU does not depend on a user-supplied value for $\pi_p$.\\
    
    \item We observe that for PN-PU classification with blob-sampling, the neural network tends to over-fit very fast causing the recall (and in some cases, also the precision) on the validation set to drop soon after the commencement of training. An obvious explanation of this phenomenon could be as follows. While high spatial variation of spectral signature is a distinctive feature of hyperspectral images, the positive samples annotated by the user happen to be localized in a small part of the image. These localized set of positive examples fail to capture the entire range of variability of the positive class in the HSI scene. This prevents the neural network from learning the right spatial invariances for the positive class causing the false negative rate shoot up. In addition to this, imperfect sampling of the negative class from $\mathcal{I}_U$ also introduces some positive samples that are labelled negative in the training set. This further contributes to the false negative rate. In order to address this problem, we stop the training as soon as the recall on the validation set starts to drop.
\end{itemize}

\section{Conclusion}
\label{sec:conclusion}
This paper takes a novel approach to HSI classification by formulating it in the PU learning paradigm. The result is a framework that is material, device and platform agnostic and can perform large scale information retrieval in arbitrary HSI scenes. We propose two approaches to solve the HSI classification problem in this framework and preliminary results on benchmark HSI datasets show promising performance. A notable drawback of the proposed approaches is the fact that every execution of the algorithms requires retraining a neural network. This poses substantial computational burden. One possible way to ameliorate this is to pre-train a neural network for a related task and retrain only the last layer for PU learning. In traditional information retrieval iterative refinement of the retrieval model based on relevance feedback plays an important role in improving the quality of retrieval from a given dataset. We plan to explore these topics in future work.\\
\begin{acks}
\label{sec:acknowledgement}
We are thankful to Zhu et al. \cite{zhu2017unsupervised} for sharing their implementation of PDHG clustering. We would also like to thank Kiryo et al \cite{kiryo2017positive} for sharing their code for the generic PU Learning framework. This study was performed as a part of the project titled "Deep Learning for Automated Feature Discovery in Hyperspectral Images (LDH)" sponsored by Space Applications Centre (SAC), Indian Space Research Organization (ISRO). Anirban Santara's work in this project was supported by Google India under the Google India Ph.D. Fellowship Award.
\end{acks}
%
\vspace{1cm}
\bibliographystyle{ACM-Reference-Format}
\bibliography{my_citations}


\begin{thebibliography}{00}


\ifx \showCODEN    \undefined \def \showCODEN     #1{\unskip}     \fi
\ifx \showDOI      \undefined \def \showDOI       #1{#1}\fi
\ifx \showISBNx    \undefined \def \showISBNx     #1{\unskip}     \fi
\ifx \showISBNxiii \undefined \def \showISBNxiii  #1{\unskip}     \fi
\ifx \showISSN     \undefined \def \showISSN      #1{\unskip}     \fi
\ifx \showLCCN     \undefined \def \showLCCN      #1{\unskip}     \fi
\ifx \shownote     \undefined \def \shownote      #1{#1}          \fi
\ifx \showarticletitle \undefined \def \showarticletitle #1{#1}   \fi
\ifx \showURL      \undefined \def \showURL       {\relax}        \fi
\providecommand\bibfield[2]{#2}
\providecommand\bibinfo[2]{#2}
\providecommand\natexlab[1]{#1}
\providecommand\showeprint[2][]{arXiv:#2}

\bibitem[\protect\citeauthoryear{Baumgardner, Biehl, and Landgrebe}{Baumgardner
  et~al\mbox{.}}{2015}]%
        {PURR1947}
\bibfield{author}{\bibinfo{person}{Marion~F. Baumgardner},
  \bibinfo{person}{Larry~L. Biehl}, {and} \bibinfo{person}{David~A.
  Landgrebe}.} \bibinfo{year}{2015}\natexlab{}.
\newblock \bibinfo{title}{220 Band AVIRIS Hyperspectral Image Data Set: June
  12, 1992 Indian Pine Test Site 3}.
\newblock   (\bibinfo{date}{Sep} \bibinfo{year}{2015}).
\newblock
\showDOI{%
\url{https://doi.org/doi:/10.4231/R7RX991C}}


\bibitem[\protect\citeauthoryear{Buchel and Ersoy}{Buchel and Ersoy}{2018}]%
        {Buchel2018LadderNF}
\bibfield{author}{\bibinfo{person}{Julian Buchel} {and}
  \bibinfo{person}{Okan~K. Ersoy}.} \bibinfo{year}{2018}\natexlab{}.
\newblock \showarticletitle{Ladder Networks for Semi-Supervised Hyperspectral
  Image Classification}.
\newblock \bibinfo{journal}{{\em CoRR\/}}  \bibinfo{volume}{abs/1812.01222}
  (\bibinfo{year}{2018}).
\newblock


\bibitem[\protect\citeauthoryear{Camps-Valls and Bruzzone}{Camps-Valls and
  Bruzzone}{2005}]%
        {Valls:2005}
\bibfield{author}{\bibinfo{person}{G. Camps-Valls} {and} \bibinfo{person}{L.
  Bruzzone}.} \bibinfo{year}{2005}\natexlab{}.
\newblock \showarticletitle{Kernel-based methods for hyperspectral image
  classification}.
\newblock \bibinfo{journal}{{\em IEEE Trans. Geosci. Remote Sens.\/}}
  \bibinfo{volume}{43}, \bibinfo{number}{6} (\bibinfo{year}{2005}),
  \bibinfo{pages}{1351--1362}.
\newblock


\bibitem[\protect\citeauthoryear{Camps-Valls, Marsheva, and Zhou}{Camps-Valls
  et~al\mbox{.}}{2007}]%
        {CampsValls2007SemiSupervisedGH}
\bibfield{author}{\bibinfo{person}{Gustavo Camps-Valls},
  \bibinfo{person}{Tatyana V.~Bandos Marsheva}, {and} \bibinfo{person}{Dengyong
  Zhou}.} \bibinfo{year}{2007}\natexlab{}.
\newblock \showarticletitle{Semi-Supervised Graph-Based Hyperspectral Image
  Classification}.
\newblock \bibinfo{journal}{{\em IEEE Transactions on Geoscience and Remote
  Sensing\/}}  \bibinfo{volume}{45} (\bibinfo{year}{2007}),
  \bibinfo{pages}{3044--3054}.
\newblock


\bibitem[\protect\citeauthoryear{Camps-Valls, Tuia, Bruzzone, and
  Benediktsson}{Camps-Valls et~al\mbox{.}}{2013}]%
        {Tuia:2013}
\bibfield{author}{\bibinfo{person}{Gustavo Camps-Valls}, \bibinfo{person}{Devis
  Tuia}, \bibinfo{person}{Lorenzo Bruzzone}, {and} \bibinfo{person}{Jón~Atli
  Benediktsson}.} \bibinfo{year}{2013}\natexlab{}.
\newblock \showarticletitle{Advances in Hyperspectral Image Classification:
  Earth monitoring with statistical learning methods}.
\newblock \bibinfo{journal}{{\em arXiv:1310.5107 [cs.CV]\/}}
  (\bibinfo{year}{2013}).
\newblock


\bibitem[\protect\citeauthoryear{Chambolle and Pock}{Chambolle and
  Pock}{2011}]%
        {chambolle2011first}
\bibfield{author}{\bibinfo{person}{Antonin Chambolle} {and}
  \bibinfo{person}{Thomas Pock}.} \bibinfo{year}{2011}\natexlab{}.
\newblock \showarticletitle{A first-order primal-dual algorithm for convex
  problems with applications to imaging}.
\newblock \bibinfo{journal}{{\em Journal of mathematical imaging and vision\/}}
  \bibinfo{volume}{40}, \bibinfo{number}{1} (\bibinfo{year}{2011}),
  \bibinfo{pages}{120--145}.
\newblock


\bibitem[\protect\citeauthoryear{Cui, Xie, Hao, Cui, and Lu}{Cui
  et~al\mbox{.}}{2018}]%
        {Cui2018SemiSupervisedCO}
\bibfield{author}{\bibinfo{person}{Binge Cui}, \bibinfo{person}{Xiaoyun Xie},
  \bibinfo{person}{Siyuan Hao}, \bibinfo{person}{Jiandi Cui}, {and}
  \bibinfo{person}{Yan Lu}.} \bibinfo{year}{2018}\natexlab{}.
\newblock \showarticletitle{Semi-Supervised Classification of Hyperspectral
  Images Based on Extended Label Propagation and Rolling Guidance Filtering}.
\newblock \bibinfo{journal}{{\em Remote Sensing\/}}  \bibinfo{volume}{10}
  (\bibinfo{year}{2018}), \bibinfo{pages}{515}.
\newblock


\bibitem[\protect\citeauthoryear{Denis, Gilleron, and Letouzey}{Denis
  et~al\mbox{.}}{2000}]%
        {Denis2000LearningFP}
\bibfield{author}{\bibinfo{person}{François Denis}, \bibinfo{person}{R{\'e}mi
  Gilleron}, {and} \bibinfo{person}{Fabien Letouzey}.}
  \bibinfo{year}{2000}\natexlab{}.
\newblock \showarticletitle{Learning from positive and unlabeled examples}.
\newblock \bibinfo{journal}{{\em Theor. Comput. Sci.\/}}  \bibinfo{volume}{348}
  (\bibinfo{year}{2000}), \bibinfo{pages}{70--83}.
\newblock


\bibitem[\protect\citeauthoryear{Du~Plessis, Niu, and Sugiyama}{Du~Plessis
  et~al\mbox{.}}{2015}]%
        {du2015convex}
\bibfield{author}{\bibinfo{person}{Marthinus Du~Plessis}, \bibinfo{person}{Gang
  Niu}, {and} \bibinfo{person}{Masashi Sugiyama}.}
  \bibinfo{year}{2015}\natexlab{}.
\newblock \showarticletitle{Convex formulation for learning from positive and
  unlabeled data}. In \bibinfo{booktitle}{{\em International Conference on
  Machine Learning}}. \bibinfo{pages}{1386--1394}.
\newblock


\bibitem[\protect\citeauthoryear{du~Plessis, Niu, and Sugiyama}{du~Plessis
  et~al\mbox{.}}{2014}]%
        {du2014analysis}
\bibfield{author}{\bibinfo{person}{Marthinus~C du Plessis},
  \bibinfo{person}{Gang Niu}, {and} \bibinfo{person}{Masashi Sugiyama}.}
  \bibinfo{year}{2014}\natexlab{}.
\newblock \showarticletitle{Analysis of learning from positive and unlabeled
  data}. In \bibinfo{booktitle}{{\em Advances in neural information processing
  systems}}. \bibinfo{pages}{703--711}.
\newblock


\bibitem[\protect\citeauthoryear{Elkan and Noto}{Elkan and Noto}{2008}]%
        {elkan2008learning}
\bibfield{author}{\bibinfo{person}{Charles Elkan} {and} \bibinfo{person}{Keith
  Noto}.} \bibinfo{year}{2008}\natexlab{}.
\newblock \showarticletitle{Learning classifiers from only positive and
  unlabeled data}. In \bibinfo{booktitle}{{\em Proceedings of the 14th ACM
  SIGKDD international conference on Knowledge discovery and data mining}}.
  ACM, \bibinfo{pages}{213--220}.
\newblock


\bibitem[\protect\citeauthoryear{F.Melgani and B.Lorenzo}{F.Melgani and
  B.Lorenzo}{2004}]%
        {Melgani:2004}
\bibfield{author}{\bibinfo{person}{F.Melgani} {and}
  \bibinfo{person}{B.Lorenzo}.} \bibinfo{year}{Aug. 2004}\natexlab{}.
\newblock \showarticletitle{Classification of hyperspectral remote sensing
  images with support vector machines}.
\newblock \bibinfo{journal}{{\em IEEE Trans. Geosci. Remote Sens.\/}}
  \bibinfo{volume}{42}, \bibinfo{number}{8} (\bibinfo{year}{Aug. 2004}),
  \bibinfo{pages}{1778--1790}.
\newblock


\bibitem[\protect\citeauthoryear{Geography}{Geography}{2018}]%
        {multi-vs-hyper}
\bibfield{author}{\bibinfo{person}{GIS Geography}.}
  \bibinfo{year}{2018}\natexlab{}.
\newblock \bibinfo{title}{Multispectral vs Hyperspectral Imagery Explained}.
\newblock   (\bibinfo{year}{2018}).
\newblock
\showURL{%
\url{https://gisgeography.com/multispectral-vs-hyperspectral-imagery-explained/}}


\bibitem[\protect\citeauthoryear{Goodfellow, Bengio, and Courville}{Goodfellow
  et~al\mbox{.}}{2016}]%
        {goodfellow2016deep}
\bibfield{author}{\bibinfo{person}{Ian Goodfellow}, \bibinfo{person}{Yoshua
  Bengio}, {and} \bibinfo{person}{Aaron Courville}.}
  \bibinfo{year}{2016}\natexlab{}.
\newblock \bibinfo{booktitle}{{\em Deep learning}}.
\newblock \bibinfo{publisher}{MIT press}.
\newblock


\bibitem[\protect\citeauthoryear{Herold, Roberts, Gardner, and Dennison}{Herold
  et~al\mbox{.}}{2004}]%
        {Herold2004SpectrometryFU}
\bibfield{author}{\bibinfo{person}{M. Herold}, \bibinfo{person}{Dar~A.
  Roberts}, \bibinfo{person}{Margaret~E. Gardner}, {and}
  \bibinfo{person}{Philip~E. Dennison}.} \bibinfo{year}{2004}\natexlab{}.
\newblock \showarticletitle{Spectrometry for urban area remote sensing —
  Development and analysis of a spectral library from 350 to 2400 nm}.
\newblock


\bibitem[\protect\citeauthoryear{Hou, Chaib-draa, Li, and Zhao}{Hou
  et~al\mbox{.}}{2018}]%
        {Hou2018GenerativeAP}
\bibfield{author}{\bibinfo{person}{Ming Hou}, \bibinfo{person}{Brahim
  Chaib-draa}, \bibinfo{person}{Chao Li}, {and} \bibinfo{person}{Qibin Zhao}.}
  \bibinfo{year}{2018}\natexlab{}.
\newblock \showarticletitle{Generative Adversarial Positive-Unlabelled
  Learning}. In \bibinfo{booktitle}{{\em IJCAI}}.
\newblock


\bibitem[\protect\citeauthoryear{Kiros, Salakhutdinov, and Zemel}{Kiros
  et~al\mbox{.}}{2014}]%
        {kiros2014unifying}
\bibfield{author}{\bibinfo{person}{Ryan Kiros}, \bibinfo{person}{Ruslan
  Salakhutdinov}, {and} \bibinfo{person}{Richard~S Zemel}.}
  \bibinfo{year}{2014}\natexlab{}.
\newblock \showarticletitle{Unifying visual-semantic embeddings with multimodal
  neural language models}.
\newblock \bibinfo{journal}{{\em arXiv preprint arXiv:1411.2539\/}}
  (\bibinfo{year}{2014}).
\newblock


\bibitem[\protect\citeauthoryear{Kiryo, Niu, du~Plessis, and Sugiyama}{Kiryo
  et~al\mbox{.}}{2017}]%
        {kiryo2017positive}
\bibfield{author}{\bibinfo{person}{Ryuichi Kiryo}, \bibinfo{person}{Gang Niu},
  \bibinfo{person}{Marthinus~C du Plessis}, {and} \bibinfo{person}{Masashi
  Sugiyama}.} \bibinfo{year}{2017}\natexlab{}.
\newblock \showarticletitle{Positive-unlabeled learning with non-negative risk
  estimator}. In \bibinfo{booktitle}{{\em Advances in Neural Information
  Processing Systems}}. \bibinfo{pages}{1675--1685}.
\newblock


\bibitem[\protect\citeauthoryear{Krizhevsky, Sutskever, and Hinton}{Krizhevsky
  et~al\mbox{.}}{2012}]%
        {Kirzhevsky:2012}
\bibfield{author}{\bibinfo{person}{Alex Krizhevsky}, \bibinfo{person}{Ilya
  Sutskever}, {and} \bibinfo{person}{Geoffrey~E. Hinton}.}
  \bibinfo{year}{2012}\natexlab{}.
\newblock \showarticletitle{ImageNet classification with deep convolutional
  neural networks}.
\newblock \bibinfo{journal}{{\em NIPS\/}} (\bibinfo{year}{2012}).
\newblock


\bibitem[\protect\citeauthoryear{Landgrebe}{Landgrebe}{2002}]%
        {Landgrebe:2002}
\bibfield{author}{\bibinfo{person}{D. Landgrebe}.}
  \bibinfo{year}{2002}\natexlab{}.
\newblock \showarticletitle{Hyperspectral image data analysis}.
\newblock \bibinfo{journal}{{\em IEEE Signal Process. Mag.\/}}
  \bibinfo{volume}{19}, \bibinfo{number}{1} (\bibinfo{year}{2002}),
  \bibinfo{pages}{17--28}.
\newblock


\bibitem[\protect\citeauthoryear{Langdon}{Langdon}{2011}]%
        {ROC_article}
\bibfield{author}{\bibinfo{person}{W.B. Langdon}.}
  \bibinfo{year}{2011}\natexlab{}.
\newblock \bibinfo{title}{Receiver Operating Characteristics (ROC)}.
\newblock   (\bibinfo{year}{2011}).
\newblock
\showURL{%
\url{http://www0.cs.ucl.ac.uk/staff/ucacbbl/roc/}}


\bibitem[\protect\citeauthoryear{Lee and Liu}{Lee and Liu}{2003}]%
        {Lee2003LearningWP}
\bibfield{author}{\bibinfo{person}{Wee~Sun Lee} {and} \bibinfo{person}{Bing
  Liu}.} \bibinfo{year}{2003}\natexlab{}.
\newblock \showarticletitle{Learning with Positive and Unlabeled Examples Using
  Weighted Logistic Regression}. In \bibinfo{booktitle}{{\em ICML}}.
\newblock


\bibitem[\protect\citeauthoryear{Li, Wu, Zhang, and Du}{Li
  et~al\mbox{.}}{2016}]%
        {Li:2016}
\bibfield{author}{\bibinfo{person}{Wei Li}, \bibinfo{person}{Guodong Wu},
  \bibinfo{person}{Fan Zhang}, {and} \bibinfo{person}{Qian Du}.}
  \bibinfo{year}{Nov. 2016}\natexlab{}.
\newblock \showarticletitle{Hyperspectral Image Classification Using Deep
  Pixel-Pair Features}.
\newblock \bibinfo{journal}{{\em IEEE Trans. Geosci. Remote Sens.\/}}
  \bibinfo{volume}{PP}, \bibinfo{number}{99} (\bibinfo{year}{Nov. 2016}),
  \bibinfo{pages}{1--10}.
\newblock
\showDOI{%
\url{https://doi.org/10.1109/TGRS.2016.2616355}}


\bibitem[\protect\citeauthoryear{Lin, Yang, Hsiao, and Chen}{Lin
  et~al\mbox{.}}{2015}]%
        {lin2015deep}
\bibfield{author}{\bibinfo{person}{Kevin Lin}, \bibinfo{person}{Huei-Fang
  Yang}, \bibinfo{person}{Jen-Hao Hsiao}, {and} \bibinfo{person}{Chu-Song
  Chen}.} \bibinfo{year}{2015}\natexlab{}.
\newblock \showarticletitle{Deep learning of binary hash codes for fast image
  retrieval}. In \bibinfo{booktitle}{{\em Proceedings of the IEEE conference on
  computer vision and pattern recognition workshops}}. \bibinfo{pages}{27--35}.
\newblock


\bibitem[\protect\citeauthoryear{Liu, Zhang, Lu, and Ma}{Liu
  et~al\mbox{.}}{2007}]%
        {liu2007survey}
\bibfield{author}{\bibinfo{person}{Ying Liu}, \bibinfo{person}{Dengsheng
  Zhang}, \bibinfo{person}{Guojun Lu}, {and} \bibinfo{person}{Wei-Ying Ma}.}
  \bibinfo{year}{2007}\natexlab{}.
\newblock \showarticletitle{A survey of content-based image retrieval with
  high-level semantics}.
\newblock \bibinfo{journal}{{\em Pattern recognition\/}} \bibinfo{volume}{40},
  \bibinfo{number}{1} (\bibinfo{year}{2007}), \bibinfo{pages}{262--282}.
\newblock


\bibitem[\protect\citeauthoryear{Mitra, Diaz, and Craswell}{Mitra
  et~al\mbox{.}}{2017}]%
        {mitra2017learning}
\bibfield{author}{\bibinfo{person}{Bhaskar Mitra}, \bibinfo{person}{Fernando
  Diaz}, {and} \bibinfo{person}{Nick Craswell}.}
  \bibinfo{year}{2017}\natexlab{}.
\newblock \showarticletitle{Learning to match using local and distributed
  representations of text for web search}. In \bibinfo{booktitle}{{\em
  Proceedings of the 26th International Conference on World Wide Web}}.
  International World Wide Web Conferences Steering Committee,
  \bibinfo{pages}{1291--1299}.
\newblock


\bibitem[\protect\citeauthoryear{Nigam, McCallum, Thrun, and Mitchell}{Nigam
  et~al\mbox{.}}{1998}]%
        {Nigam1998LearningTC}
\bibfield{author}{\bibinfo{person}{Kamal Nigam}, \bibinfo{person}{Andrew
  McCallum}, \bibinfo{person}{Sebastian Thrun}, {and}
  \bibinfo{person}{Tom~Michael Mitchell}.} \bibinfo{year}{1998}\natexlab{}.
\newblock \showarticletitle{Learning to Classify Text from Labeled and
  Unlabeled Documents}. In \bibinfo{booktitle}{{\em AAAI/IAAI}}.
\newblock


\bibitem[\protect\citeauthoryear{Powers}{Powers}{2011}]%
        {powers2011evaluation}
\bibfield{author}{\bibinfo{person}{David~Martin Powers}.}
  \bibinfo{year}{2011}\natexlab{}.
\newblock \showarticletitle{Evaluation: from precision, recall and F-measure to
  ROC, informedness, markedness and correlation}.
\newblock  (\bibinfo{year}{2011}).
\newblock


\bibitem[\protect\citeauthoryear{Rao, Garg, and Ghosh}{Rao
  et~al\mbox{.}}{2007}]%
        {Rao2007DevelopmentOA}
\bibfield{author}{\bibinfo{person}{Nallani Venkata~Rama Rao},
  \bibinfo{person}{P.~K. Garg}, {and} \bibinfo{person}{Sunil~Kumar Ghosh}.}
  \bibinfo{year}{2007}\natexlab{}.
\newblock \showarticletitle{Development of an agricultural crops spectral
  library and classification of crops at cultivar level using hyperspectral
  data}.
\newblock \bibinfo{journal}{{\em Precision Agriculture\/}}  \bibinfo{volume}{8}
  (\bibinfo{year}{2007}), \bibinfo{pages}{173--185}.
\newblock


\bibitem[\protect\citeauthoryear{Richards}{Richards}{2013}]%
        {Richards:2013}
\bibfield{author}{\bibinfo{person}{J.A. Richards}.}
  \bibinfo{year}{2013}\natexlab{}.
\newblock \bibinfo{booktitle}{{\em Remote Sensing Digital Image Analysis: An
  Introduction}}.
\newblock \bibinfo{publisher}{Springer, New York, NY, USA}.
\newblock


\bibitem[\protect\citeauthoryear{Santara, Mani, Hatwar, Singh, Garg, Padia, and
  Mitra}{Santara et~al\mbox{.}}{2017}]%
        {santara2017bass}
\bibfield{author}{\bibinfo{person}{Anirban Santara}, \bibinfo{person}{Kaustubh
  Mani}, \bibinfo{person}{Pranoot Hatwar}, \bibinfo{person}{Ankit Singh},
  \bibinfo{person}{Ankur Garg}, \bibinfo{person}{Kirti Padia}, {and}
  \bibinfo{person}{Pabitra Mitra}.} \bibinfo{year}{2017}\natexlab{}.
\newblock \showarticletitle{BASS Net: band-adaptive spectral-spatial feature
  learning neural network for hyperspectral image classification}.
\newblock \bibinfo{journal}{{\em IEEE Transactions on Geoscience and Remote
  Sensing\/}} \bibinfo{volume}{55}, \bibinfo{number}{9} (\bibinfo{year}{2017}),
  \bibinfo{pages}{5293--5301}.
\newblock


\bibitem[\protect\citeauthoryear{Schmidhuber}{Schmidhuber}{2015}]%
        {schmidhuber2015deep}
\bibfield{author}{\bibinfo{person}{J{\"u}rgen Schmidhuber}.}
  \bibinfo{year}{2015}\natexlab{}.
\newblock \showarticletitle{Deep learning in neural networks: An overview}.
\newblock \bibinfo{journal}{{\em Neural networks\/}}  \bibinfo{volume}{61}
  (\bibinfo{year}{2015}), \bibinfo{pages}{85--117}.
\newblock


\bibitem[\protect\citeauthoryear{Shepherd and Walsh}{Shepherd and Walsh}{[n.
  d.]}]%
        {ShepherdDevelopmentOR}
\bibfield{author}{\bibinfo{person}{Keith~D. Shepherd} {and}
  \bibinfo{person}{Markus~G. Walsh}.} \bibinfo{year}{[n. d.]}\natexlab{}.
\newblock \showarticletitle{Development of Reflectance Spectral Libraries for
  Characterization of Soil Properties}.
\newblock


\bibitem[\protect\citeauthoryear{Smeulders, Worring, Santini, Gupta, and
  Jain}{Smeulders et~al\mbox{.}}{2000}]%
        {smeulders2000content}
\bibfield{author}{\bibinfo{person}{Arnold~WM Smeulders},
  \bibinfo{person}{Marcel Worring}, \bibinfo{person}{Simone Santini},
  \bibinfo{person}{Amarnath Gupta}, {and} \bibinfo{person}{Ramesh Jain}.}
  \bibinfo{year}{2000}\natexlab{}.
\newblock \showarticletitle{Content-based image retrieval at the end of the
  early years}.
\newblock \bibinfo{journal}{{\em IEEE Transactions on Pattern Analysis \&
  Machine Intelligence\/}} \bibinfo{number}{12} (\bibinfo{year}{2000}),
  \bibinfo{pages}{1349--1380}.
\newblock


\bibitem[\protect\citeauthoryear{Tokui, Oono, Hido, and Clayton}{Tokui
  et~al\mbox{.}}{2015}]%
        {tokui2015chainer}
\bibfield{author}{\bibinfo{person}{Seiya Tokui}, \bibinfo{person}{Kenta Oono},
  \bibinfo{person}{Shohei Hido}, {and} \bibinfo{person}{Justin Clayton}.}
  \bibinfo{year}{2015}\natexlab{}.
\newblock \showarticletitle{Chainer: a next-generation open source framework
  for deep learning}. In \bibinfo{booktitle}{{\em Proceedings of workshop on
  machine learning systems (LearningSys) in the twenty-ninth annual conference
  on neural information processing systems (NIPS)}}, Vol.~\bibinfo{volume}{5}.
  \bibinfo{pages}{1--6}.
\newblock


\bibitem[\protect\citeauthoryear{Van~den Oord, Dieleman, and Schrauwen}{Van~den
  Oord et~al\mbox{.}}{2013}]%
        {van2013deep}
\bibfield{author}{\bibinfo{person}{Aaron Van~den Oord}, \bibinfo{person}{Sander
  Dieleman}, {and} \bibinfo{person}{Benjamin Schrauwen}.}
  \bibinfo{year}{2013}\natexlab{}.
\newblock \showarticletitle{Deep content-based music recommendation}. In
  \bibinfo{booktitle}{{\em Advances in neural information processing systems}}.
  \bibinfo{pages}{2643--2651}.
\newblock


\bibitem[\protect\citeauthoryear{Wan, Wang, Hoi, Wu, Zhu, Zhang, and Li}{Wan
  et~al\mbox{.}}{2014}]%
        {wan2014deep}
\bibfield{author}{\bibinfo{person}{Ji Wan}, \bibinfo{person}{Dayong Wang},
  \bibinfo{person}{Steven Chu~Hong Hoi}, \bibinfo{person}{Pengcheng Wu},
  \bibinfo{person}{Jianke Zhu}, \bibinfo{person}{Yongdong Zhang}, {and}
  \bibinfo{person}{Jintao Li}.} \bibinfo{year}{2014}\natexlab{}.
\newblock \showarticletitle{Deep learning for content-based image retrieval: A
  comprehensive study}. In \bibinfo{booktitle}{{\em Proceedings of the 22nd ACM
  international conference on Multimedia}}. ACM, \bibinfo{pages}{157--166}.
\newblock


\bibitem[\protect\citeauthoryear{Wang, Yang, Ooi, Zhang, and Zhuang}{Wang
  et~al\mbox{.}}{2016}]%
        {wang2016effective}
\bibfield{author}{\bibinfo{person}{Wei Wang}, \bibinfo{person}{Xiaoyan Yang},
  \bibinfo{person}{Beng~Chin Ooi}, \bibinfo{person}{Dongxiang Zhang}, {and}
  \bibinfo{person}{Yueting Zhuang}.} \bibinfo{year}{2016}\natexlab{}.
\newblock \showarticletitle{Effective deep learning-based multi-modal
  retrieval}.
\newblock \bibinfo{journal}{{\em The VLDB Journal—The International Journal
  on Very Large Data Bases\/}} \bibinfo{volume}{25}, \bibinfo{number}{1}
  (\bibinfo{year}{2016}), \bibinfo{pages}{79--101}.
\newblock


\bibitem[\protect\citeauthoryear{Yoshitaka and Ichikawa}{Yoshitaka and
  Ichikawa}{1999}]%
        {yoshitaka1999survey}
\bibfield{author}{\bibinfo{person}{Atsuo Yoshitaka} {and}
  \bibinfo{person}{Tadao Ichikawa}.} \bibinfo{year}{1999}\natexlab{}.
\newblock \showarticletitle{A survey on content-based retrieval for multimedia
  databases}.
\newblock \bibinfo{journal}{{\em IEEE Transactions on Knowledge and Data
  Engineering\/}} \bibinfo{volume}{11}, \bibinfo{number}{1}
  (\bibinfo{year}{1999}), \bibinfo{pages}{81--93}.
\newblock


\bibitem[\protect\citeauthoryear{Zhang and Zuo}{Zhang and Zuo}{2008}]%
        {Zhang2008LearningFP}
\bibfield{author}{\bibinfo{person}{Bangzuo Zhang} {and} \bibinfo{person}{Wanli
  Zuo}.} \bibinfo{year}{2008}\natexlab{}.
\newblock \showarticletitle{Learning from Positive and Unlabeled Examples: A
  Survey}.
\newblock \bibinfo{journal}{{\em 2008 International Symposiums on Information
  Processing\/}} (\bibinfo{year}{2008}), \bibinfo{pages}{650--654}.
\newblock


\bibitem[\protect\citeauthoryear{Zhu, Chayes, Tiard, Sanchez, Dahlberg,
  Bertozzi, Osher, Zosso, and Kuang}{Zhu et~al\mbox{.}}{2017a}]%
        {zhu2017unsupervised}
\bibfield{author}{\bibinfo{person}{Wei Zhu}, \bibinfo{person}{Victoria Chayes},
  \bibinfo{person}{Alexandre Tiard}, \bibinfo{person}{Stephanie Sanchez},
  \bibinfo{person}{Devin Dahlberg}, \bibinfo{person}{Andrea~L Bertozzi},
  \bibinfo{person}{Stanley Osher}, \bibinfo{person}{Dominique Zosso}, {and}
  \bibinfo{person}{Da Kuang}.} \bibinfo{year}{2017}\natexlab{a}.
\newblock \showarticletitle{Unsupervised classification in hyperspectral
  imagery with nonlocal total variation and primal-dual hybrid gradient
  algorithm}.
\newblock \bibinfo{journal}{{\em IEEE Transactions on Geoscience and Remote
  Sensing\/}} \bibinfo{volume}{55}, \bibinfo{number}{5} (\bibinfo{year}{2017}),
  \bibinfo{pages}{2786--2798}.
\newblock


\bibitem[\protect\citeauthoryear{Zhu, Tuia, Mou, Xia, Zhang, Xu, and
  Fraundorfer}{Zhu et~al\mbox{.}}{2017b}]%
        {zhu2017deep}
\bibfield{author}{\bibinfo{person}{Xiao~Xiang Zhu}, \bibinfo{person}{Devis
  Tuia}, \bibinfo{person}{Lichao Mou}, \bibinfo{person}{Gui-Song Xia},
  \bibinfo{person}{Liangpei Zhang}, \bibinfo{person}{Feng Xu}, {and}
  \bibinfo{person}{Friedrich Fraundorfer}.} \bibinfo{year}{2017}\natexlab{b}.
\newblock \showarticletitle{Deep learning in remote sensing: A comprehensive
  review and list of resources}.
\newblock \bibinfo{journal}{{\em IEEE Geoscience and Remote Sensing
  Magazine\/}} \bibinfo{volume}{5}, \bibinfo{number}{4} (\bibinfo{year}{2017}),
  \bibinfo{pages}{8--36}.
\newblock


\end{thebibliography}

\end{document}